\documentclass{PoS}
\newcommand{\pt}{\ensuremath{p_{T}}}

\newcommand{\zg}{\ensuremath{z_g}~}
\newcommand{\rg}{\ensuremath{R_g}~}
\usepackage{url}

\title{Measurements of the jet internal sub-structure and its relevance to parton shower evolution in p+p and Au+Au collisions at STAR}

\ShortTitle{Jet sub-structure at STAR}

\author{\speaker{Raghav Kunnawalkam Elayavalli for the STAR Collaboration}\\
        Wayne State University, Detroit MI 48201\\
        E-mail: \email{raghavke@wayne.edu}}

\abstract{
Recent measurements of jet structure modifications at RHIC and LHC highlight the importance of differential measurements to study the nature of jet quenching. Since these jet structure observables are intimately dependent on parton evolution in both the angular and energy scales, measurements are needed to disentangle these two scales in order to probe the medium at different length scales to study its characteristic properties such as the coherence length. To that effect, the STAR collaboration presents fully unfolded results of a jet's sub-structure via the SoftDrop shared momentum fraction ($z_{g}$) and the groomed jet radius ($R_{g}$) in p+p collisions at $\sqrt{s} = $ 200 GeV as a function of jet momenta. Having established the p+p baseline, we present the first measurement of the jet's inherent angular structure in Au+Au collisions at $\sqrt{s_{NN}} = $ 200 GeV via an experimentally robust observable related to the SoftDrop $R_{g}$: the opening angle between the two leading sub-jets ($\theta_{SJ}$). In Au+Au collisions at STAR, we utilize a specific di-jet selection as introduced in our previous momentum imbalance ($A_{J}$) measurement and the recoil jet spectra differentially as a function of jet transverse momentum belonging to particular angular lasses based on the $\theta_{SJ}$ observable. With such measurements, we probe the medium response to jets at a particular resolution scale and find no significant differences in quenching for jets of different angular scales as given by $\theta_{SJ}$.}

\FullConference{International Conference on Hard and Electromagnetic Probes of High-Energy Nuclear Collisions\\
		30 September - 5 October 2018\\
		Aix-Les-Bains, Savoie, France}

\begin{document}

\section{Introduction}
Relativistic ion collisions produce copious amounts of jets due to the hard scatterings between quarks and gluons of the colliding nuclei. 
Recent measurements at both RHIC and LHC along with theoretical advancements have shown the importance of studying and measuring the properties of these jets both in p+p and in heavy ion collisions (reviews of jet studies can be found here~\cite{expreview}). 
There are two natural scales that characterize a jet and its evolution: the momentum and the angular scales. First generation measurements have measured jet quenching in an inclusive manner via the momentum asymmetry in di-jet events and have further extensively studied the momentum dependence via nuclear modification factors and fragmentation functions. Jet-medium interaction could further be dependent on the resolution scale or the coherence length of the medium which sees the jet as a singular radiating object or a multi-prong object~\cite{cohlength}. 

\section{Jet sub-structure in p+p Collisions}

In order to differentially study energy loss in the medium, the jet structure in vacuum has to be first understood i.e., the DGLAP splitting functions that govern parton evolution. In recent literature, SoftDrop~\cite{SoftDrop} has gained fame as an algorithm which can extract such scales experimentally with its procedure of walking backwards in the Cambridge/Aachen clustering tree until two sub-jets satisfy $z = \frac{min(p_{T,1}, p_{T,2})}{p_{T,1} + p_{T,2}} > z_{cut} (\Delta R/R)^{\beta}$ where \zg and \rg are the $z$ and $\Delta R$ upon termination of the algorithm with $z_{cut} = 0.1$ and $\beta = 0$. It was shown that for such choices of $z_{cut}, \beta$ the SoftDrop $z_{g}$ distribution converges to the vacuum DGLAP splitting functions for $z>z_{cut}$ in a ``Sudakov-safe'' manner~\cite{Larkoski:2015lea}. 

\begin{figure}[h] 
   \centering
   \includegraphics[width=0.49\textwidth]{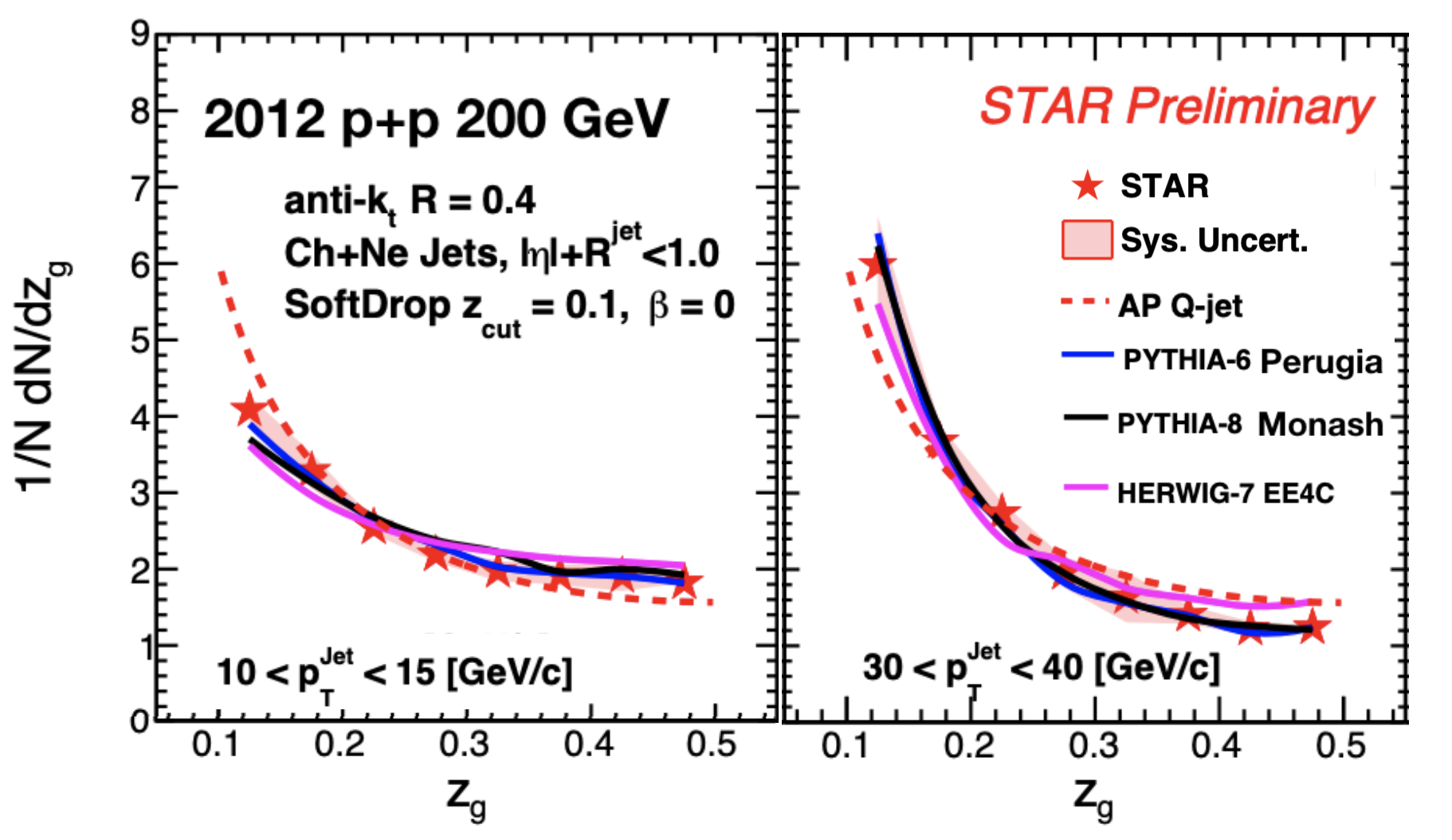} 
   \includegraphics[width=0.47\textwidth]{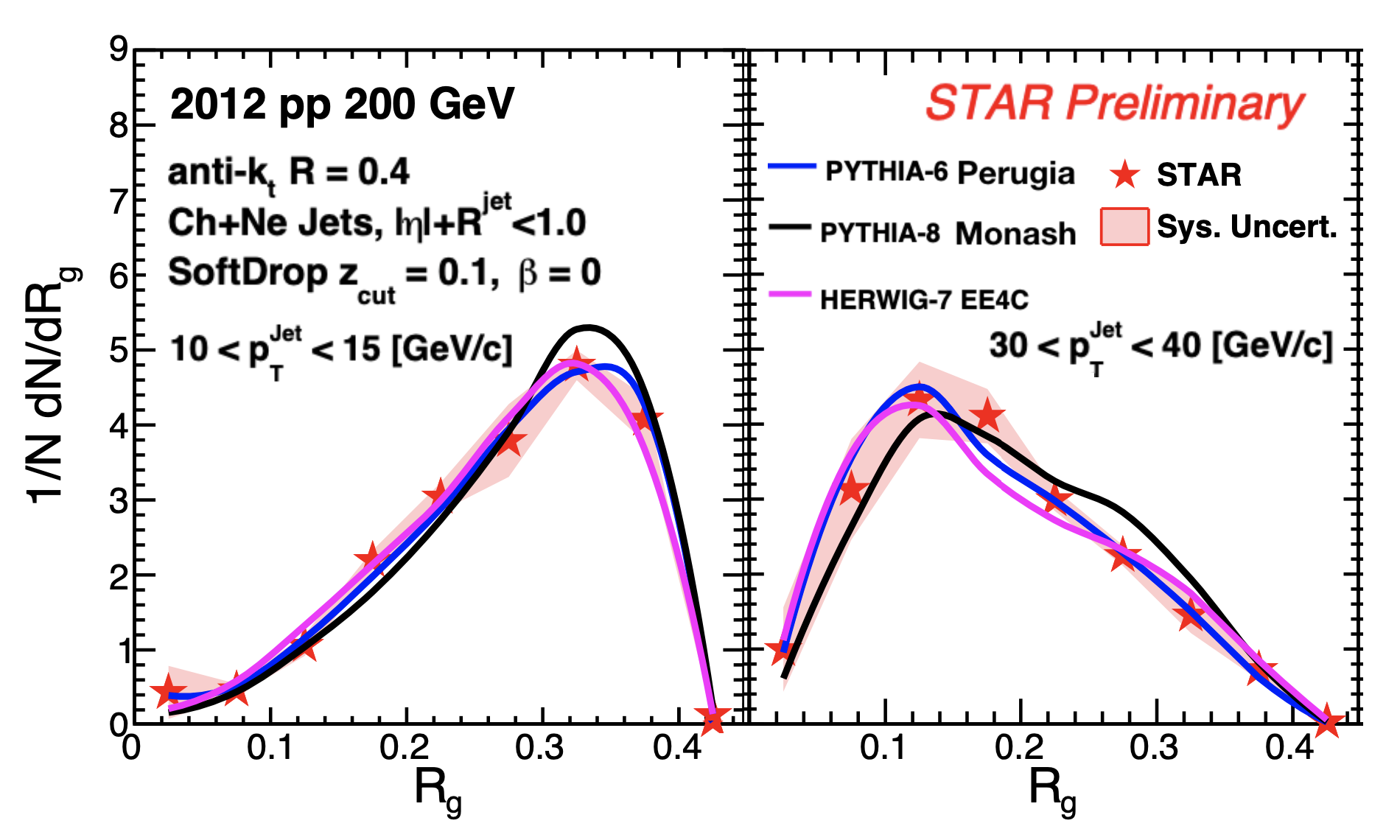} 
   \caption{Fully unfolded measurement of the SoftDrop groomed sub-jet shared momentum fraction ($z_{g}$) and the groomed jet radius ($R_{g}$) in p+p collisions at $\sqrt{s} =$ 200 GeV. The markers and lines are described in the text.}
   \label{fig:ppzgrg}
\end{figure}

The p+p data for the groomed jet structure measurements were collected with the STAR detector~\cite{star} during the 2012 run at $\sqrt{s_{NN}} = 200$ GeV.  Jets are reconstructed from charged tracks in the Time Projection Chamber (TPC) and energy depositions in the Barrel ElectroMagnetic Calorimeter (BEMC) using the anti-$k_{t}$ algorithm as implemented in the FastJet package~\cite{FastJet}, hereafter referenced as Ch+Ne jets. For additional details regarding event/track/tower quality selections, please refer to~\cite{starprl, nicktalk}. For the inclusive p+p analysis, events are selected by an online jet patch trigger which is a $1\times1$ patch in $\eta-\phi$ with the total sum $E_{T}$ in the patch to be greater than 7.3 GeV.  
Two-dimensional unfolding in $p_T$ and \zg or $R_{g}$ respectively, was done using Bayesian unfolding as implemented in the RooUnfold package~\cite{roounfold, dgostini} with four iterations. The response matrix is created from a PYTHIA-6 (Perugia Tune, slightly adapted to STAR data) prior and a GEANT-3 simulation of the STAR detector. The systematic uncertainties for data are taken as a quadrature sum resulting from the following sources: tracking efficiency ($4\%$), tower gain calibration ($3.8\%$), hadronic correction to the tower energy scale (described in ~\cite{starprl}) and unfolding related sources including varying the iteration parameter from 2 to 6 and the prior in the response matrix.  

Figure~\ref{fig:ppzgrg} shows the fully unfolded \zg and \rg distributions for two jet \pt~selections, 10-15 and 30-40 GeV/$c$, respectively. The STAR data are shown in the red filled star markers with the red shaded region corresponding to the overall systematic uncertainty. Leading order Monte Carlo (MC) generators such as PYTHIA-6 Perugia tune, PYTHIA-8 Monash, and Herwig-7 EE4C UE tune in the blue, black and magenta lines are also plotted for comparison to the data. For the \zg observable, we also provide the symmetrized DGLAP splitting functions (noted as AP Q-Jet in the figure) at leading order in the red dashed lines for quark jets (with the splitting being similar for quark and gluon initiated jets). All the models studied reproduce the general trends seen in the data, particularly the dependence on the jet momenta leading to a steeper \zg distribution and a narrower $R_{g}$. 

\section{Jet Angular Scale in Au+Au Collisions}

The Au+Au data used in these proceedings were collected during the 2007 run with its corresponding reference p+p run in 2006 at $\sqrt{s_{NN}} = 200$ GeV. Since the jet patch trigger is saturated in an Au+Au event, we employ a high tower (HT) trigger, requiring at least one BEMC tower with $E_{T}$ > 5.4 GeV. Event centrality in Au+Au is determined by the raw charged track multiplicity in the TPC within $|\eta| < 0.5$ and we show only events in the 0-20\% centrality range. In Au+Au events, we have two separate jet collections given by the HardCore selection~\cite{starprl}, where jets are clustered with objects (tracks/towers) with $p_{T}> 2$ GeV/$c$, and Matched jets which are clustered from the constituent-subtracted~\cite{cs} event with our nominal $p_{T} > 0.2$ GeV/$c$ for constituents, and are geometrically matched to the HardCore jets ($\Delta R < 0.4$). Further event selection criteria include a minimum \pt~requirement for HardCore di-jets ($p^{\rm{Lead}}_{T} > 16, p^{\rm{SubLead}}_{T} > 8$ GeV/$c$) and an azimuthal angle ($|\Delta \phi (\rm{Lead, SubLead})| > 2\pi/3$) selection to focus on back-to-back di-jets. 

\begin{figure}[h] 
   \centering
   \includegraphics[width=0.7\textwidth]{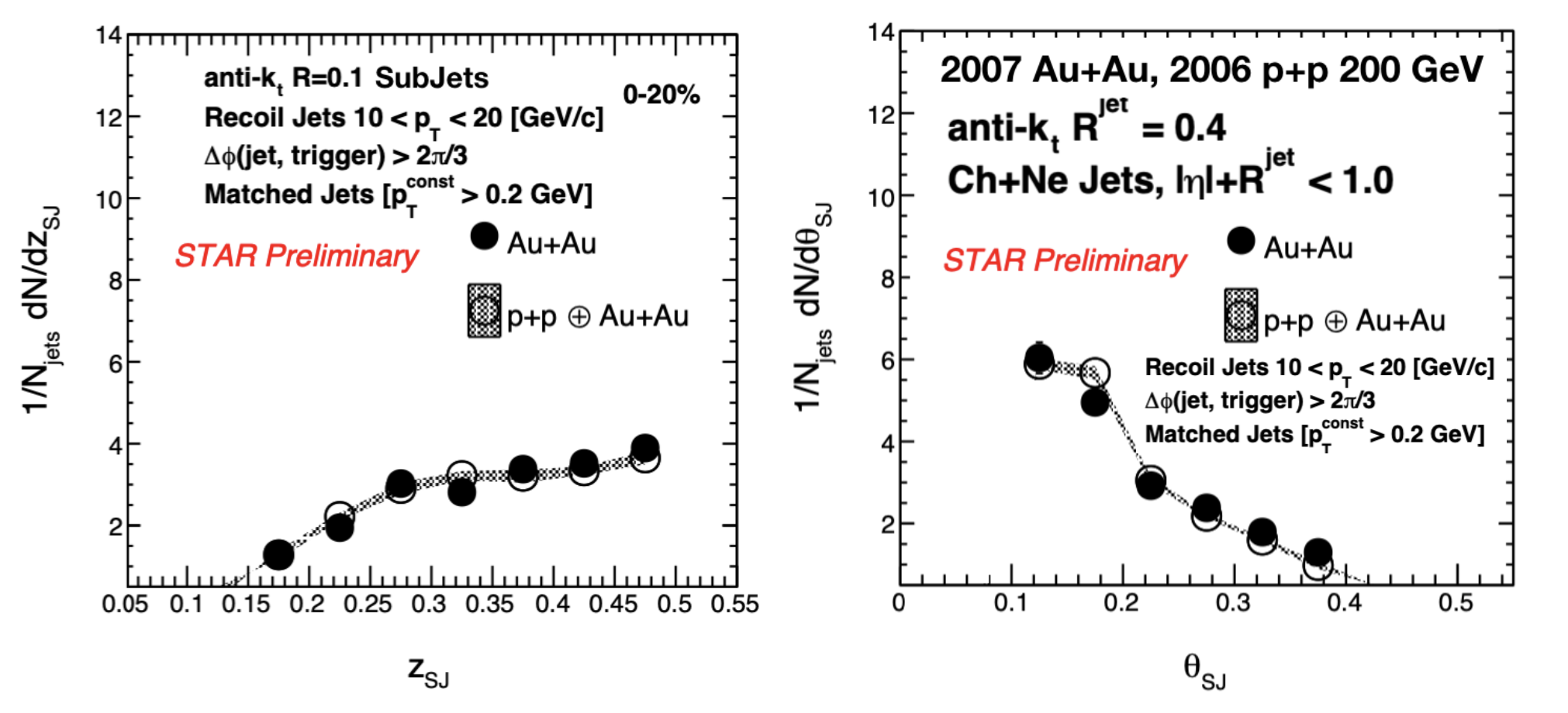} 
   \caption{TwoSubJet observables ($z_{SJ}$ - left and $\theta_{SJ}$ - right) comparing Au+Au (solid circles) and p+p embedded in Au+Au (open circles including the systematic uncertainties in the shaded region) for recoil Matched jets (anti-$k_{t}$ $R=0.4$ jets with $R=0.1$ sub-jets) in the $p_{T}$ range 10-20 GeV$/c$.  }
   \label{fig:twoSubJet}
\end{figure}

In our studies, we found the groomed jet radii (R$_{g}$) to be highly sensitive to the fluctuating underlying event found in Au+Au collisions and therefore we devised a new observable involving sub-jets of a smaller radius reconstructed within the original jet (see here~\cite{SubJetliliana} for a recent theoretical article demonstrating similar classes of observables).
For our nominal anti-$k_{t}$ jets of $R=0.4$, we reconstruct an inclusive set of anti-$k_{t}$ sub-jets with $R=0.1$ from the original jet's constituents. An absolute minimum sub-jet \pt~requirement of $2.97$ GeV/$c$ is enforced in central Au+Au collisions to reduce sensitivity to the background fluctuations. The two observables related to the momentum and angular scales are then defined as follows  $z_{SJ} = \frac{min(p^{SJ1}_{T}, p^{SJ2}_{T})}{p^{SJ1}_{T} + p^{SJ2}_{T}}, ~ ~ ~ \theta_{SJ} = \Delta R (SJ1, SJ2) $
where $SJ1, SJ2$ are the leading and sub-leading sub-jets, respectively. 

\begin{figure}[h] 
   \centering
   \includegraphics[width=0.55\textwidth]{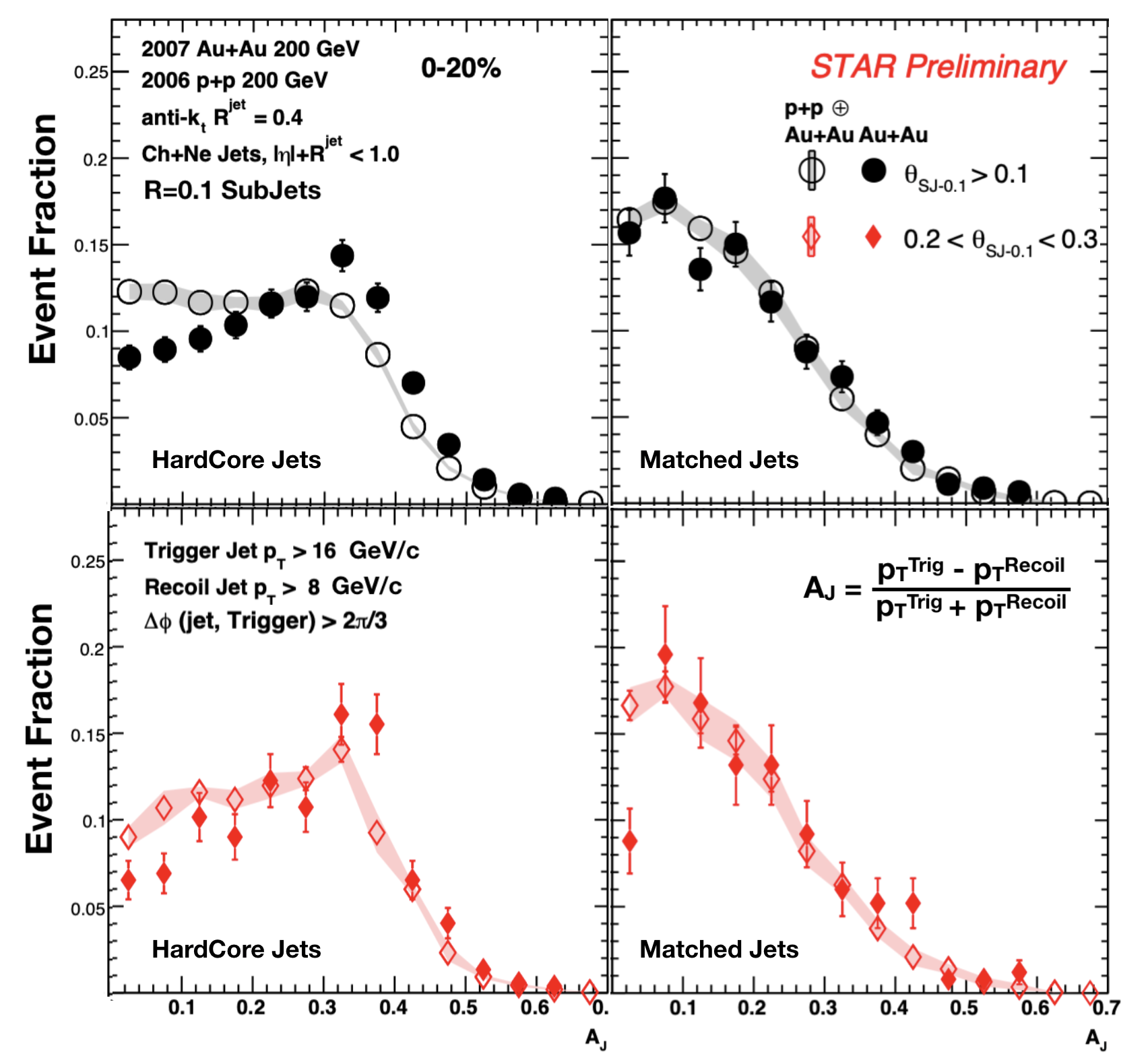} 
   \includegraphics[width=0.38\textwidth]{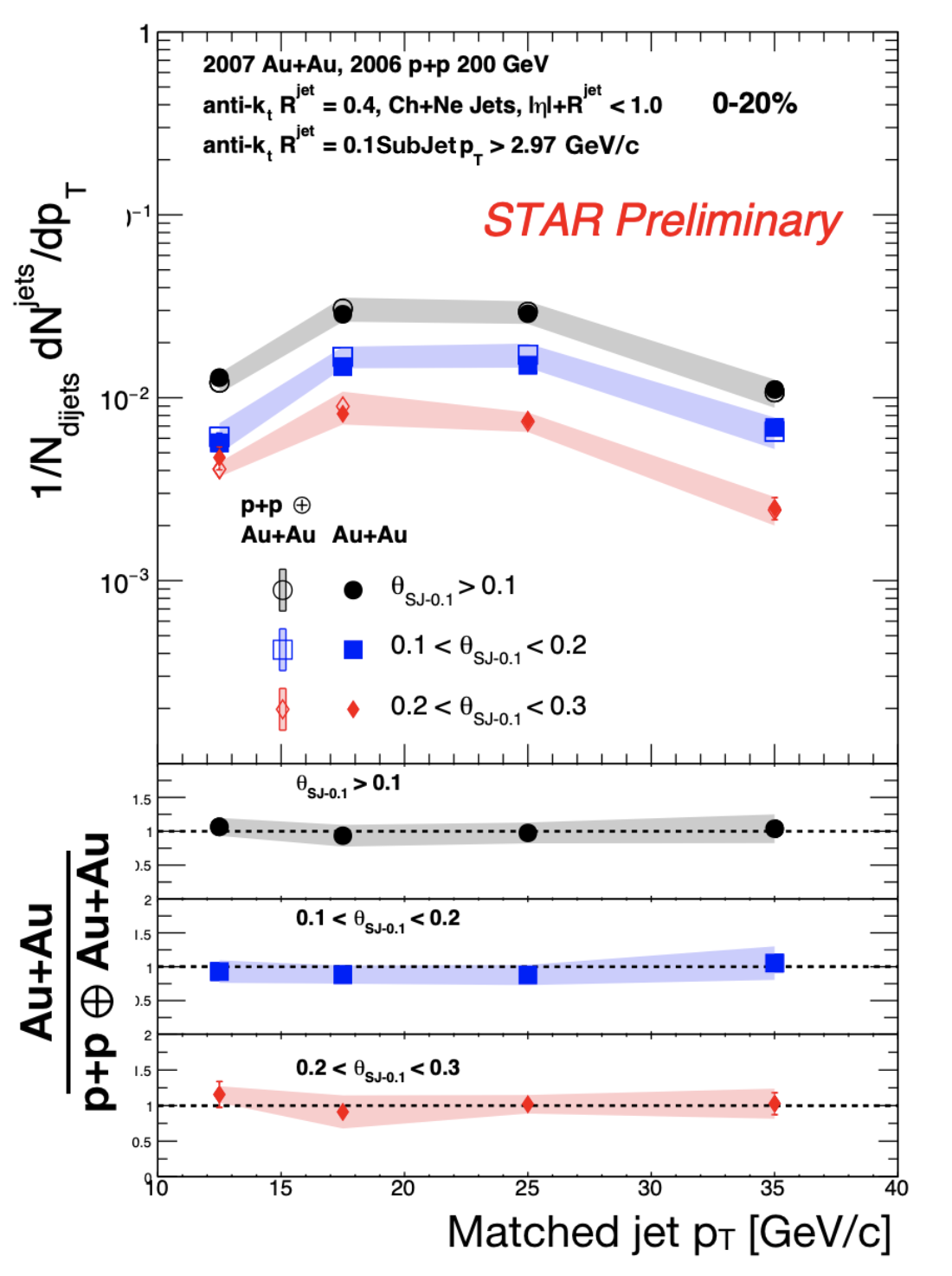} 
   \caption{HardCore and Matched (left figure) di-jet asymmetry ($|A_{J}|$) and the Matched recoil jet yield (right figure) along with the ratios shown in the bottom right panels. Markers are described in the text. 
   }
   \label{fig:aj}
\end{figure}

For a meaningful comparison between Au+Au and a p+p reference, the effects of background fluctuations and detector inefficiencies must be taken into account. To achieve this, HT-triggered p+p data from 2006 is embedded into minimum bias Au+Au data (p+p $\oplus$ Au+Au) from 2007, in the same centrality range (0-20\%). During embedding, we account for the relative tracking efficiency (90\%$\pm$7\%) and relative tower energy scale (100\%$\pm$2\%), with a one sigma variation taken as systematic uncertainties.
Figure~\ref{fig:twoSubJet} shows the detector level TwoSubJet $z_{SJ}$ (left) and $\theta_{SJ}$ (right) distributions for constituent-subtracted Matched recoil jets with $R=0.1$ sub-jets (henceforth denoted as $SJ-0.1$ in the figures) recoiling off the trigger (selected with a $|\Delta \phi(\rm{jet, HT})| > 2\pi/3$), in the \pt~range 10-20 GeV/$c$. Results are observed to be similar in both Au+Au and p+p $\oplus$ Au+Au. We also observe a remarkable difference in the shape of $z_{SJ}$ when compared to that of the SoftDrop $z_{g}$, 
which is caused by selecting the core of the jet. The $\theta_{SJ}$ for jets within the considered $p_{T}$ range peaks at small values and includes a natural lower cutoff at the sub-jet radius and we now select jets based on this distribution. 

Di-jet asymmetry for both HardCore and Matched jets (left panels) are shown in Figure~\ref{fig:aj}. The right panels of Figure~\ref{fig:aj} show the yield of Matched recoil jets normalized per di-jet for the different $\theta_{SJ}$ selections and the ratios of Au+Au/p+p in the bottom panels. The black, blue and red markers represent recoil jets with selections on $\theta_{SJ}$ $[0.1, 0.4], [0.1, 0.2]$ and $[0.2, 0.3]$, respectively, for inclusive, narrow and wide jets. We observe a clear di-jet imbalance indicating jet quenching effects in the $|A_{J}|$ distributions (comparing Au+Au to p+p$\oplus$Au+Au) for all HardCore jets including the wide angle jets. The Matched jets on the other hand are balanced at RHIC energies, as evident by ratios in the bottom right panels consistent with unity. This is consistent with our earlier measurement~\cite{starprl}. We also note that wide angle jets are still balanced indicating no apparent distinction between wide and narrow jets by the medium in our selection. Further detailed differential analyses are required with the high statistics 2014 data set to extract the medium resolution scale or the coherence length and the effect on standard jet quenching observables at RHIC energies. 


\section{Conclusions}
STAR has presented the first fully unfolded SoftDrop \zg and \rg measurements of inclusive jets with varying transverse momentum in p+p collisions at $\sqrt{s} = $200 GeV. The measurements are overall reproduced by current leading order Monte Carlo event generators for both \zg and \rg for jets in our kinematic acceptance and reflect the momentum dependent narrowing of jet structure. Due to the sensitivity of the SoftDrop observables to the Au+Au underlying event, we introduce and measure the TwoSubJet observables, $z_{SJ}$ and $\theta_{SJ}$ for R=0.1 anti-$k_{t}$ sub-jets as representing the momentum and angular scales of a jet in a heavy ion environment. We measure the di-jet momentum asymmetry and the recoil jet yield with the special di-jet selection at STAR and find that HardCore di-jets are imbalanced and Matched di-jets are balanced for jets of varying angular scales. We find no significant difference in the quenching phenomenon for both wide and narrow jets leading to the conclusion that these special jets do not undergo significantly different jet-medium interactions due to their varying angular scales.

\end{document}